\RequirePackage{ifpdf}
\ifpdf 
\documentclass[pdftex]{sigma}
\else
\documentclass{sigma}
\fi

\begin{document}
\allowdisplaybreaks

\newcommand{\semi}{\subset \hskip -4.2mm +}
\newcommand{\gen}[1]{\partial_{#1}}
\newcommand{\curl}[1]{ \{#1\} }

\renewcommand{\PaperNumber}{014}

\FirstPageHeading

\ShortArticleName{On the Virasoro Structure of Symmetry Algebras}

\ArticleName{On the Virasoro Structure of Symmetry Algebras\\ of
Nonlinear Partial Dif\/ferential Equations}

\Author{Faruk G\"UNG\"OR}
\AuthorNameForHeading{F. G\"ung\"or}

\Address{Department of Mathematics, Faculty of Science and
Letters, Istanbul Technical University,\\
34469, Istanbul, Turkey}

\Email{\href{mailto:gungorf@itu.edu.tr}{gungorf@itu.edu.tr}}
\URLaddress{\url{http://www.mat.itu.edu.tr/gungor/}}

\ArticleDates{Received November 30, 2005, in f\/inal form January 20,
2006; Published online January 30, 2006}

\Abstract{We discuss  Lie algebras of the Lie symmetry groups of two
generically non-integrable equations in one temporal and two space
dimensions arising in dif\/ferent contexts. The f\/irst is a
generalization of the KP equation and contains 9~arbitrary functions
of one and two arguments. The second one is a system of PDEs that
depend on some physical parameters. We require that these PDEs are
invariant under a Kac--Moody--Virasoro algebra. This leads to several
limitations on the coef\/f\/icients (either functions or parameters)
under which equations are prime candidates for being integrable.}


\Classification{35A30; 35Q53; 35Q55; 35Q58}

\section{Introduction}

It is well known that a number of physically signif\/icant integrable
partial dif\/ferential equations in 2+1-dimensions  typically have
inf\/inite-dimensional symmetry algebras with a specif\/ic
Kac--Moody--Virasoro (KMV) structure. Among them one can cite the
Kadomtsev--Petviashvili (KP) equation \cite{David85, David86},
modif\/ied KP, cylindrical KP equation \cite{Levi88}, all equations of
the KP hierarchy~\cite{Orlov97}, Davey--Stewartson (DS) system
\cite{Champagne88} and three-wave resonant interaction equations
\cite{Martina89}. On the other hand,  it should be mentioned that
there are  evolution type equations which are integrable, but do
admit inf\/inite-dimensional symmetry algebras without a KMV
structure. For instance, the breaking soliton equation and
Zakharov--Strachan equation \cite{Velan98} do not allow a KMV type
symmetry algebra while they are integrable. This observation shows
that the existence of a KMV symmetry algebra is not a necessary
condition for integrability for a nonlinear evolution equation in
2+1 dimensions. Nevertheless, it is  our f\/irm belief that
identifying  equations with KMV symmetry algebra can serve us to
provide  those subclasses which are candidates for integrability. Of
course, integrable ones must be further singled out  by checking
them for integrability in any sense of the word. To mention a few,
one can perform singularity analysis for establishing Painlev\'e
property or proceed to study higher symmetries or look for a Lax
pair.

In this paper we shall concentrate on generalizations of two
integrable equations in two space dimensions. One is the generalized
KP (GKP) equation (prototype of a 2+1-dimensional integrable
equation) and the other is generalized DS (GDS) equation. For both
systems we shall determine the cases when the equations admit an
inf\/inite-dimensional symmetry group the Lie algebra of which has a
Virasoro structure. It will be shown that how this requirement will
impose  restrictions on the coef\/f\/icients.

First we consider the GKP equation
\begin{gather}
 (u_t+p(t)uu_{x}+q(t)u_{xxx})_{x}+\sigma(y,t)u_{yy}+a(y,t)u_{y} \nonumber\\
\qquad{}+ b(y,t)u_{xy}+c(y,t)u_{xx}+e(y,t)u_{x}+f(y,t)u+h(y,t)=0.\label{1.1}
\end{gather}
We assume that in some neighbourhood we have
\begin{gather*}
p(t)\ne 0,\qquad q(t)\ne 0,\qquad \sigma(y,t)\ne 0.
\end{gather*}
The other functions in \eqref{1.1} are arbitrary.

Second, we consider the system
\begin{gather}
i\psi_t+\delta \psi_{xx}+\psi_{yy}=\chi |\psi|^2\psi+\gamma
(w_x+\phi_y)\psi,\nonumber\\
w_{xx}+n\phi_{xy}+m_2w_{yy}=(|\psi|^2)_{x},\nonumber\\
nw_{xy}+\lambda\phi_{xx}+m_1\phi_{yy}=(|\psi|^2)_{y},\label{GDS0}
\end{gather}
with the condition $(\lambda-1)(m_1-m_2)=n^2$. Here $\psi(t,x,y)$
is a complex function, $w(t,x,y)$ and $\phi(t,x,y)$ are real
functions and $\delta$, $n$, $m_1$, $m_2$, $\lambda$, $\chi$, $\gamma$ are
real constants. This system of nonlinear partial dif\/ferential
equations in 2+1 dimensions arises as a model of wave propagation
in a bulk medium composed of an elastic material with couple
stresses \cite{Babaoglu04}.

\section[Kac-Moody-Virasoro (KMV) algebras and their subalgebras]{Kac--Moody--Virasoro (KMV)
algebras and their subalgebras}

We recall that a  KMV algebra is an inf\/inite-dimensional Lie
algebra with a basis \cite{Winternitz88, Winternitz89}
\begin{gather}\label{KMV}
\{L_m, T_m^a, C, K\},\qquad 1\leq a \leq N,\quad m\in \mathbb{Z},
\end{gather}
satisfying the commutation relations
\begin{gather}
[L_m, L_n]=(m-n)L_{m+n}+\frac{1}{12}m\left(m^2-1\right)\delta_{m,-n}C,\nonumber\\
[T_m^a, T_m^b]=f^{abc}T_{m+n}^c+m\delta_{a,b}\delta_{m,-n}K,\qquad
[L_m, T_n^a]=-nT_{m+n}^a,\nonumber \\
[C, L_m]=[C, T_m^a]=[K, L_m]=[K, T_m^a]=[K, C]=0,\label{comm}
\end{gather}
where $f^{abc}$ are the structure constants of some
f\/inite-dimensional real or complex simple Lie algebra $A$. The
elements $\curl{T_m^a, K}$ form the basis of a Kac--Moody algebra,
$\curl{L_m, C}$ form the Virasoro algebra, $K$ and $C$ are central
elements, namely they commute with all other elements and with each
other. From~\eqref{comm}, it is seen that the Kac--Moody algebra is
an ideal in the entire structure. KMV algebras seem to arise in many
branches of theoretical physics and in the theory of completely
integrable systems.

A simple realization of the algebra   \eqref{KMV} is obtained by introducing a scalar parameter $\lambda$,
 a~f\/inite-dimensional Lie algebra $A$ with basis $\curl{X^1,\ldots, X^N}$ and commutation relations
\begin{gather*}
[X^a, X^b]=f^{abc}X^c,\qquad     a, b, c=1,2,\ldots, N.
\end{gather*}
We put
\begin{gather}\label{realiz}
L_m=-\lambda^{m+1}\gen   \lambda,\qquad T_m^a=X^a\lambda^m,\qquad
C=0,\qquad K=0
\end{gather}
and see that the commutation relations \eqref{comm} are satisf\/ied.
As the central elements $C$ and $K$ are represented trivially in
\eqref{realiz}, we actually have a representation of an af\/f\/ine
loop algebra. For example, the set of (complex) vector f\/ields on
the unit circle
\[
L_m=-ie^{imt}\gen t
\]
 satisf\/ies
\[
[L_m, L_n]=(m-n)L_{m+n},
\]
 and therefore
realizes a Virasoro algebra without a center.

\subsection{The symmetry algebra of CGKP equation}

With the introduction of allowed transformations  \cite{Gungor02-2} (point
transformations taking equation into one with dif\/ferent
coef\/f\/icients) we can transform \eqref{1.1} into some canonical
form
\begin{gather}
(u_t+uu_x+u_{xxx})_x+\varepsilon u_{yy}+a(y,t)u_y+b(y,t)u_{xy}\nonumber\\
\qquad{}+ c(y,t)u_{xx}+f(y,t)u=0,\qquad  \varepsilon=\pm 1.\label{canon}
\end{gather}
We call \eqref{canon} canonical GKP equation and abbreviate CGKP
equation.

If we restrict ourselves to Lie point symmetries of equation
\eqref{canon}, the Lie algebra of the symmetry group is
represented by vector f\/ields of the form
\begin{gather}\label{3.1}
{\boldsymbol{V}}=\xi\gen x+\eta\gen y+\tau\gen t+\phi\gen u,
\end{gather}
where $\xi$, $\eta$, $\tau$ and $\phi$ are functions of $x$, $y$, $t$
and $u$. The method for the determination of the coef\/f\/icients of
the vector f\/ield $\boldsymbol{V}$ is algorithmic \cite{Olver91}.
Applying this to \eqref{canon} gives  an overdetermined set of
linear partial dif\/ferential equations for the coef\/f\/icients $\xi$,
$\eta$, $\tau$ and $\phi$ in equation~\eqref{3.1}. Solving this system
we f\/ind that the vector f\/ield \eqref{3.1} should have the form
\begin{gather}
{\boldsymbol{V}}=\tau(t)\gen t+\left(\frac{1} {3}\dot \tau x +
\xi _0 (y,t)\right)\gen x\nonumber\\
\phantom{{\mathbf{V}}=}{}+\Bigl(\frac{2} {3}\dot \tau y + \eta _0 (t)\Bigr)\gen
y+\left(-\frac{2} {3}\dot \tau u + \frac{1} {3}\ddot \tau x +
S(y,t)\right)\gen u,\label{3.2}
\end{gather}
where
\begin{gather}\label{3.3}
S(y,t) =  - \tau c_t  - \left(\frac{2} {3}\dot \tau y + \eta _0 \right)c_y +
\xi _{0,t}  + b\xi _{0,y}  - \frac{2} {3}c\dot \tau,
\end{gather}
and $\tau(t)$, $\eta(t)$ and $\xi_0(y,t)$ satisfy
\begin{gather}
3\tau a_t  + (2\dot \tau y + 3\eta _0 )a_y  + 2a\dot \tau  = 0,\nonumber 
\\
 - 3\dot \eta _0  - 2y\ddot \tau  + 3\tau b_t  + (2\dot \tau y + 3\eta _0 )b_y
  + b\dot \tau  - 6\varepsilon \xi _{0,y}  = 0, \nonumber \label{3.5}
\\
\ddot \tau  + 3a\xi _{0,y}  + 3\varepsilon \xi _{0,yy}  = 0, \nonumber 
\\
f\ddot \tau  = 0, \nonumber
\\
4f\dot \tau  + 3f_t \tau  + f_y (2\dot \tau y + 3\eta _0 ) = 0, \nonumber
\\
\dddot \tau  + 3fS + 3aS_y  + 3\varepsilon S_{yy}  = 0,
\label{3.9}
 \end{gather}
where $S(y,t)$ of equation \eqref{3.3} should be substituted into equation~\eqref{3.9}.

We shall determine the conditions on the functions $a$, $b$, $c$
and $f$ that allow the symmetry algebra to be inf\/inite-dimensional
with an additional  KMV structure.

\subsection{Virasoro symmetries of the CGKP
equation}

The canonical generalized KP equation  \eqref{canon} will be invariant
under a transformation group, the Lie algebra of which is
isomorphic to a Virasoro algebra if the function $\tau$ in
\eqref{3.2} remains free. Omitting details we summarize the main
results as a theorem:

\begin{theorem}\label{t1}
The canonical generalized KP equation \eqref{canon} allows the Virasoro algebra as
a symmetry algebra if and only if the coef\/f\/icients satisfy
\begin{gather}\label{4.10}
a=f=0,\quad b=b(t),\qquad c=c_0(t)+c_1(t)y.
\end{gather}
\end{theorem}

\begin{theorem}\label{t2}
The canonical generalized KP equation
\begin{gather}\label{4.11}
(u_t+uu_x+u_{xxx})_x+\varepsilon
u_{yy}+b(t)u_{xy}+[c_0(t)+c_1(t)y]u_{xx}=0
\end{gather}
with $\varepsilon=\pm 1$, and $b(t)$, $c_0(t)$ and $c_1(t)$
arbitrary smooth functions is invariant under an
inf\/inite-dimensional Lie point symmetry group. Its Lie algebra has
a Kac--Moody--Virasoro structure. It is realized by vector f\/ields of
the form
\begin{gather}\label{4.12}
{\boldsymbol{V}}=T(\tau)+X(\xi)+Y(\eta),
\end{gather}
where $\tau(t)$, $\xi(t)$ and $\eta(t)$ are arbitrary smooth
functions of time and we have
\begin{gather}
  T(\tau ) = \tau (t)\partial _t  + \frac{1}
{6}\big[3\varepsilon \dot by\tau  + (2x + \varepsilon by)\dot \tau  - \varepsilon \ddot \tau y^2 \big]\partial _x
   + \frac{2}{3}\dot \tau \partial _y \nonumber\\
\phantom{T(\tau ) =}{}
   + \frac{1}
{6}\big\{ \big[ - 6\dot c_0  + 3\varepsilon b\dot b + ( - 6\dot c_1  + 3\varepsilon \ddot b)y\big]\tau
   + \big[ - 4u + \varepsilon b^2  - 4c_0  + 4(\varepsilon \dot b - 2c_1 )y\big]\dot \tau \nonumber\\
\phantom{T(\tau ) =}{}
   + (2x - \varepsilon by)\ddot \tau     - \varepsilon y^2 \dddot \tau \big\} \partial _u,  \label{4.13}
\\
   X(\xi ) = \xi (t)\partial _x  + \dot \xi (t)\partial _u,  \label{4.14} \\
  Y(\eta ) = \eta (t)\partial _y  - \frac{\varepsilon }
{2}\dot \eta (t)y\partial _x  - \frac{1} {2}[2c_1 \eta
   + \varepsilon b\dot \eta  + \varepsilon y\ddot \eta ]\partial
   _u.\label{4.15}
\end{gather}
\end{theorem}

\begin{remark}
The transformation
\begin{gather}
  u(x,y,t) = \tilde u(\tilde x,\tilde y,\tilde t) + \left(\frac{\varepsilon }
{2}\dot b - c_1 \right)y - c_0  + \frac{\varepsilon }
{4}b^2,  \nonumber \\
  \tilde x = x - \frac{{\varepsilon b}}
{2}y,\qquad \tilde y = y,\qquad \tilde t = t \label{4.16}
\end{gather}
takes equation \eqref{4.11} into the KP equation itself, i.e.\ into equation
\eqref{4.11} with $b=c_0=c_1=0$. The transformation \eqref{4.16}
also transforms the Lie algebra \eqref{4.12}--\eqref{4.15} into
the symmetry algebra~\cite{David85} of the KP equation.
\end{remark}

\begin{theorem}\label{t3}
The GKP equation \eqref{1.1} is invariant under a Lie point
symmetry group, the Lie algebra of which contains a Virasoro
algebra as a subalgebra, if and only if it can be transformed into
the KP equation itself by a point transformation.
\end{theorem}

In summary, we have identif\/ied all cases when the generalized KP
equation has an inf\/inite-dimensional symmetry group whose Lie
algebra has a Virasoro structure. For a complete analysis and some
implications of these results we refer the reader to~\cite{Gungor02-2}.

A natural question to ask is what can one do with the integrable
CGKP \eqref{4.11}?  The tools of soliton theory such as the inverse
spectral transform and B\"acklund transformations are at our
disposal  to obtain multisoliton and other physically important
solutions of this equation.

A general class of fourth order scalar partial dif\/ferential
equations, invariant under the group of local point transformations
of the KP equation has been constructed in \cite{David88}. We note
that a~recent paper \cite{Lou04} has been devoted to the
construction of equations of arbitrary order invariant under the KP
symmetry group. They searched for an autonomous higher order KP
family
\[
\left(u_t+\frac{3}{2}uu_x+u_{xxx}\right)_x+\frac{3}{4}\delta u_{yy}+F(u)=0,
\]
where $F(u)$ is a  dif\/ferential function of $u$ and its any order
derivatives of $x$, $y$, $t$ variables with no space-time dependence
explicitly occurring in it, which possesses the same KMV symmetry
algebra as the standard KP equation. The idea used in~\cite{Lou04}
is similar to the one we used above. The KP family is required to
be left invariant under the KP symmetry group. The Virasoro
algebra will be present in the entire symmetry algebra as a
subalgebra whenever symmetry equations (obtained from splitting of
the linearized equation) can be solved identically with the
function $\tau(t)$ ($t$-coef\/f\/icient of the vector f\/ield)
arbitrary. This implies that the equation is invariant under an
arbitrary reparametrization of time. These type of equations will
include the most probable candidates for integrability (integrable
models).

In another paper \cite{Gungor04-1}, an attempt was made to
determine whether there can exist any possible variable
coef\/f\/icient extensions of the KP equation
\begin{gather*}
\bigl(u_t+f(x,y,t)uu_x+g(x,y,t)u_{xxx}\bigr)_x+h(x,y,t)u_{yy}=0,
\end{gather*}
such that it can not be transformed to the standard KP equation by
allowed transformations, but still can have an
inf\/inite-dimensional symmetry group.

We now turn to the system \eqref{GDS0}. For our purposes we f\/ind
it  more convenient to study the dif\/ferentiated form of it with
the f\/irst complex equation separated into real and imaginary
components via $\psi=u+iv$
\begin{gather}
u_t+\delta v_{xx}+v_{yy}=\chi v\left(u^2+v^2\right)+\gamma v(w+\phi),\nonumber\\
-v_t+\delta u_{xx}+u_{yy}=\chi u\left(u^2+v^2\right)+\gamma u(w+\phi),\nonumber\\
w_{xx}+n\phi_{xx}+m_2w_{yy}=2\left(u_x^2+uu_{xx}+v_x^2+vv_{xx}\right),\nonumber\\
nw_{yy}+\lambda\phi_{xx}+m_1\phi_{yy}=2\left(u_y^2+uu_{yy}+v_y^2+vv_{yy}\right).\label{GDS}
\end{gather}

Again, we represent a general element of the symmetry algebra by a
vector f\/ield of the form
\begin{gather*}
    \boldsymbol{V}=\tau\gen t+\xi\gen x+\eta\gen y+\varphi_1\gen
u+\varphi_2\gen v+\varphi_3\gen w+\varphi_4\gen \phi,
\end{gather*}
where the coef\/f\/icients $\tau$, $\xi$, $\eta$, $\varphi_i$, $i=1,2,3,4$ are
functions of $t$, $x$, $y$, $u$, $v$, $w$, $\phi$. Applying the symmetry
procedure we f\/ind
\begin{gather}\label{symVF}
   \boldsymbol{V}=T(f)+X(g)+Y(h)+W(m),
\end{gather}
where
\begin{gather}
T(f)=f(t)\gen t+\frac{1}{2}f'(t)(x\gen x+y\gen y-u\gen u-v\gen
v-2w\gen w-2\phi\gen \phi)\nonumber\\
\phantom{T(f)=}{}-\frac{(x^2+\delta y^2)}{8\delta}\left[f''(t)(v\gen u-u\gen v)
+\frac{f'''(t)}{2\gamma}(\gen w+\gen \phi)\right],
\nonumber\\
X(g)=g(t)\gen x-\frac{x}{2\delta}\left[g'(t)(v\gen u-u\gen v)+\frac{g''(t)}{2\gamma}(\gen w+\gen \phi)\right],\nonumber\\
Y(h)=h(t)\gen y-\frac{y}{2}\left[h'(t)(v\gen u-u\gen v)+\frac{h''(t)}{2\gamma}(\gen w+\gen \phi)\right],\nonumber\\
W(m)=m(t)(v\gen u-u\gen v)+\frac{m'(t)}{2\gamma}(\gen w+\gen \phi).\label{comp}
\end{gather}
The functions $g(t)$, $h(t)$, and $m(t)$ are arbitrary functions
of class $C^{\infty}(I)$, $I\subseteq \mathbb{R}$. The function~$f(t)$ is arbitrary if
\begin{gather}\label{constraint}
   m_2 \delta  + n + 1 = 0, \qquad m_1 \delta + n\delta  + \lambda=0,
\end{gather}
otherwise $f(t)=c_2 t^2  + c_1 t + c_0.$ We  focus on the case
when $f(t)$ is allowed to be arbitrary for which the symmetry
algebra realized by the vector f\/ields \eqref{symVF} and
\eqref{comp} is then inf\/inite-dimensional. The commutation
relations for the GDS algebra are as follows:
\begin{gather*}
[T(f_1),T(f_2)]=T(f_1f'_2-f'_1f_2),\qquad
[T(f),X(g)]=X\left(fg'-\frac{1}{2}f'g\right),\nonumber\\
[T(f),Y(h)]=Y\left(fh'-\frac{1}{2}f'h\right),\qquad
[T(f),W(m)]=W(fm'),\nonumber\\
[X(g_1),X(g_2)]=-\frac{1}{2\delta}W(g_1g'_2-g'_1g_2),\qquad
[Y(h_1),Y(h_2)]=-\frac{1}{2}W(h_1h'_2-h'_1h_2),\nonumber\\
[X(g),Y(h)]=[X(g),W(m)]=[Y(h),W(m)]=[W(m_1),W(m_2)]=0.
\end{gather*}
They characterize the commutation relations of a centerless KMV
algebra which is identif\/ied as the semi-direct sum  (actually a
Levi decomposition)
$L=S\semi N$,
where $S=\curl{T(f)}$ is a simple inf\/inite dimensional Lie algebra
and
$
N=\curl{X(g), Y(h), W(m)}
$ is a nilpotent ideal
(nilradical).

More interestingly, the GDS \eqref{GDS} system has a Lie symmetry
algebra $L$ isomorphic to that of the DS symmetry algebra (the
symmetry algebra of the integrable DS equations)
\cite{Champagne88}
\begin{gather*}
i\psi_t+\delta_1 \psi_{xx}+\psi_{yy}=\delta_2|\psi|^2\psi+w\psi,\nonumber \\
\varepsilon_1w_{xx}+w_{yy}=\varepsilon_2(|\psi|^2)_{yy},
\end{gather*}
with $\delta_1=\pm 1$, $\delta_2=\pm 1$ and
$\delta_1+\varepsilon_1=0$.

We have the following (full details  can be found in
\cite{Gungor05-1}):

\begin{theorem}\label{t4}
The system \eqref{GDS} is invariant under an inf\/inite-dimensional
Lie point symmetry group, the Lie algebra of which has a
Kac--Moody--Virasoro structure isomorphic to the DS algebra if and
only if the conditions \eqref{constraint} hold.
\end{theorem}

Presence of  isomorphism suggests to look for a point
transformation taking symmetry algebras into each other. Hence we
can possibly expect that   the corresponding equations are
transformed into each other under such a transformation. Let us
mention that there is indeed a transformation $q=w+\phi-|\psi|^2$
reducing the GDS symmetry algebra to the DS algebra. However, this
transformation does not appear to be one-to-one because both
systems do not have the same number of  dependent variables.

\section{Conclusions}

In this paper, we studied group theoretical properties of  two
classes of nonlinear evolution equations in (2+1)-dimensions, namely
generalized KP and DS equations. In particular, we demanded these
equations to be invariant under the inf\/inite-dimensional  Lie groups
of local point transformations with additional KMV structure. We
showed that for the generalized KP equation~\eqref{1.1}, this is
possible only if the coef\/f\/icients satisfy~\eqref{4.10}. This is the
only case when the original equation can be transformed to the KP
equation itself. For the generalized DS equations, we showed that
the KMV symmetry algebra can be present only if the conditions~\eqref{constraint},
under which the symmetry algebra is isomorphic
to that of the integrable DS equations, hold.

Despite the usefulness of existence of KMV symmetry algebras for
equations in two space dimensions, many fundamental questions still
remain open. The precise relationship between symmetries and
integrability is still far from being completely understood. As  was
noted in~\cite{Velan98},  not all integrable equations admit KMV
symmetry algebras. On the other hand, all known non-integrable
equations are invariant under f\/inite or inf\/inite-dimensional Lie
point transformation groups without  KMV structure. These facts
demonstrate that a fuller understanding of the absence of KMV
structure in an inf\/inite-dimensional symmetry algebra is essential
in this context. If the existence of a KMV symmetry algebra alone
can not be a necessary condition for integrability for a nonlinear
evolution equation in 2+1 dimensions, so what conditions must be
added to ensure integrability?

\subsection*{Acknowledgements} The author is grateful to the referees for useful comments and
acknowledges the f\/inancial support from Istanbul Technical
University (ITU).

\LastPageEnding

\end{document}